\documentclass{iau_FM}
\usepackage{graphicx}
\usepackage{natbib}

\title[The Distribution and Excitation of Cometary CH$_3$OH] 
{Measuring the Distribution and Excitation of Cometary CH$_3$OH Using ALMA}

\author[Cordiner et al.]   
{M. A. Cordiner$^1$, S. B. Charnley$^1$, M. J. Mumma$^1$, D. Bockel{\'e}e-Morvan$^2$, N. Biver$^2$, G. Villanueva$^1$, L. Paganini$^1$, S. N. Milam$^1$, A. J. Remijan$^3$, D. C. Lis$^6$, J. Crovisier$^2$, J. Boissier$^4$, Y.-J. Kuan$^5$, I. M. Coulson$^7$}

\affiliation{$^1$NASA Goddard Space Flight Center, 8800 Greenbelt Road, Greenbelt, MD 20771, USA.\\ email: {\tt martin.cordiner@nasa.gov} \\
$^2$LESIA, Observatoire de Paris, CNRS, UPMC, Universit{\'e} Paris-Diderot, 
Meudon, France. \\
$^3$National Radio Astronomy Observatory, Charlottesville, VA 22903, USA. \\
$^4$IRAM, 300 Rue de la Piscine, 38406 Saint Martin d'Heres, France. \\
$^5$National Taiwan Normal University, Taipei 116, Taiwan, ROC. \\
$^6$LERMA, Observatoire de Paris, PSL Research University, CNRS, Sorbonne Universit{\'e}s, UPMC Univ. Paris 06, F-75014, Paris, France.\\
$^7$East Asian Observatory, Hilo 96720, USA.}

\pubyear{2015}
\jname{Astronomy in Focus, Volume 1} 
\editors{Piero Benvenuti, ed.}
\begin{document}

\maketitle

\begin{abstract}
The Atacama Large Millimeter/submillimeter Array (ALMA) was used to obtain measurements of spatially and spectrally resolved CH$_3$OH emission from comet C/2012 K1 (PanSTARRS) on 28-29 June 2014. Detection of 12-14 emission lines of CH$_3$OH  on each day permitted the derivation of spatially-resolved rotational temperature profiles (averaged along the line of sight), for the innermost 5000~km of the coma.  On each day, the CH$_3$OH distribution was centrally peaked and approximately consistent with spherically symmetric, uniform outflow.  The azimuthally-averaged CH$_3$OH rotational temperature ($T_{rot}$) as a function of sky-projected nucleocentric distance ($\rho$), fell by about 40~K between $\rho=0$ and 2500~km on 28 June, whereas on 29 June, $T_{rot}$ fell by about 50~K between $\rho=$~0~km and 1500~km. A remarkable ($\sim50$~K) rise in $T_{rot}$  at $\rho=$1500-2500~km on 29 June was not present on 28 June. The observed variations in CH$_3$OH rotational temperature are interpreted primarily as a result of variations in the coma kinetic temperature due to adiabatic cooling, and heating through Solar irradiation, but collisional and radiative non-LTE excitation processes also play a role.

\end{abstract}

\firstsection 

\section{Introduction}

Comets are believed to have formed around 4.5~Gyr ago and contain ice, dust and debris left over from the formation of the Solar System. Their compositions can therefore reveal information on the physical and chemical properties of the protosolar disk and prior interstellar cloud. Much of our knowledge on cometary compositions comes from remote (ground-based) observations, for which relatively low angular resolution and spatial coverage limits the amount of information that can be obtained. In particular, there is a lack of understanding concerning the physical and chemical structure of the near-nucleus coma, at distances less than a few thousand kilometres from the nucleus.

The first cometary observations with ALMA were reported by \citet{cor14}, who measured the distributions of HCN, HNC and H$_2$CO in the inner comae of comets C/2012 F6 (Lemmon) and C/2012 S1 (ISON). By virtue of its large abundance in comets and its complex energy level structure, methanol (CH$_3$OH), is the most readily-detectable molecule for probing the coma temperature at radio/sub-mm wavelengths. Here, we briefly summarise some early results that exploit the unprecedented resolution and sensitivity of ALMA to provide new information on the distribution and temperature of CH$_3$OH in the inner coma of comet C/2012 K1 (PanSTARRS). 

\section{Observations}

ALMA observations of C/2012 K1 (PanSTARRS) were obtained on 2014-06-28 19:07-20:05 and 2014-06-28 17:37-18:26 at a heliocentric distance $r_H=1.43$~AU and geocentric distance $\Delta=1.97$~AU. The CH$_3$OH $K=3-2$ band near 251.9~GHz was observed on 28 June and the $J=7-6$ band near 338.5~GHz was observed on 29 June. Atmospheric conditions were outstanding throughout (with zenith PWV~$<0.4$~mm). Thirty 12-m antennae, with baseline lengths 20-650~m,  resulted in an angular resolution of $0.80''\times0.43''$ at 252~GHz and $0.71''\times0.33''$ at 338~GHz, with 488~kHz spectral resolution. The data were flagged, calibrated and imaged using standard CASA routines \citep[see][]{cor14}. Twelve individual CH$_3$OH emission lines (originating from the $K=3$ rotational level, with $\Delta{K}=-1$, $\Delta{J}=0$) were detected on 28 June and fourteen lines (originating from the $J=7$ level, with $\Delta{J}=-1$, $\Delta{K}=0$) were detected on 29 June.

\begin{figure*}
\centering
\includegraphics[width=0.45\textwidth]{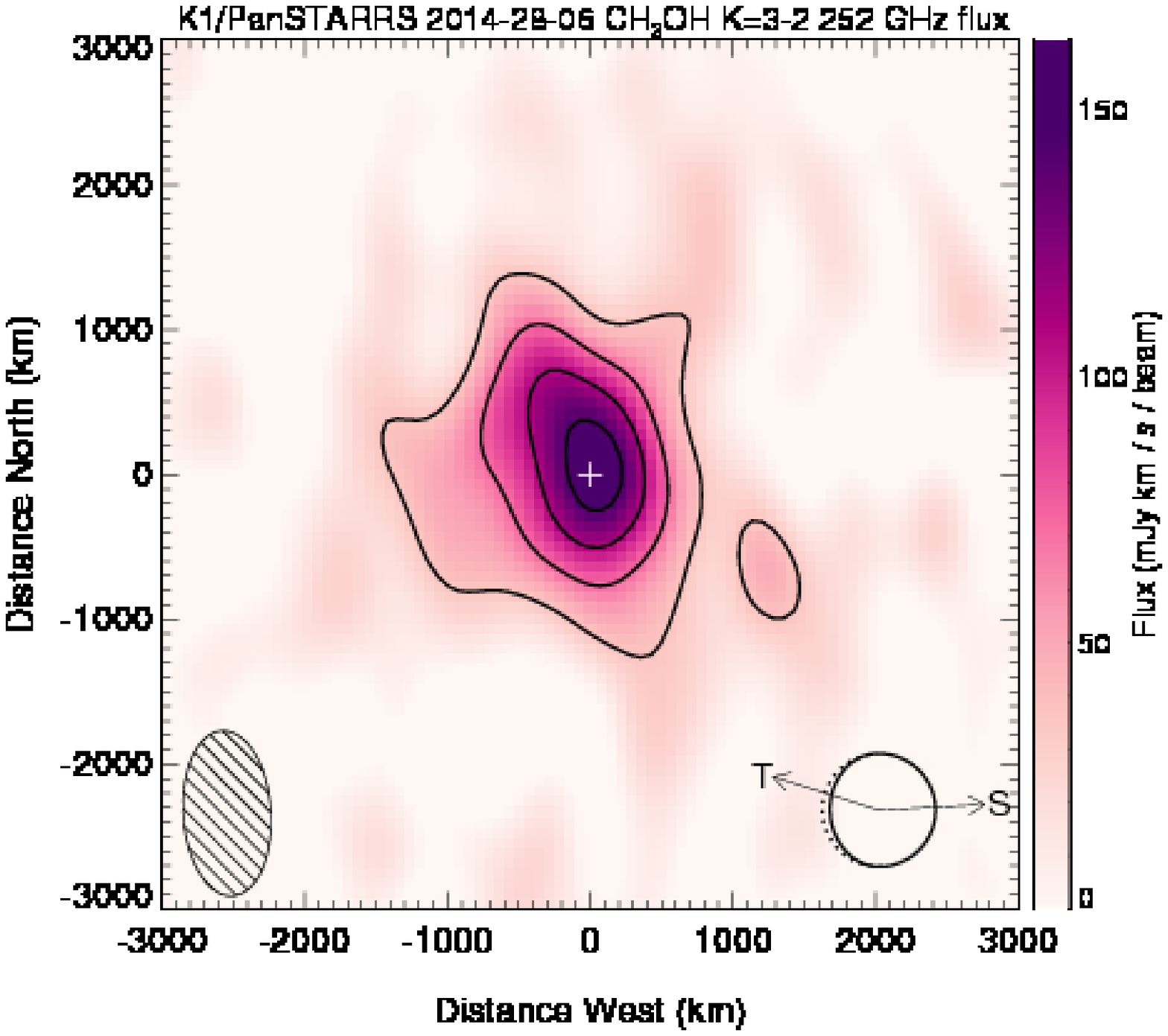}
\hspace{2mm}
\includegraphics[width=0.45\textwidth]{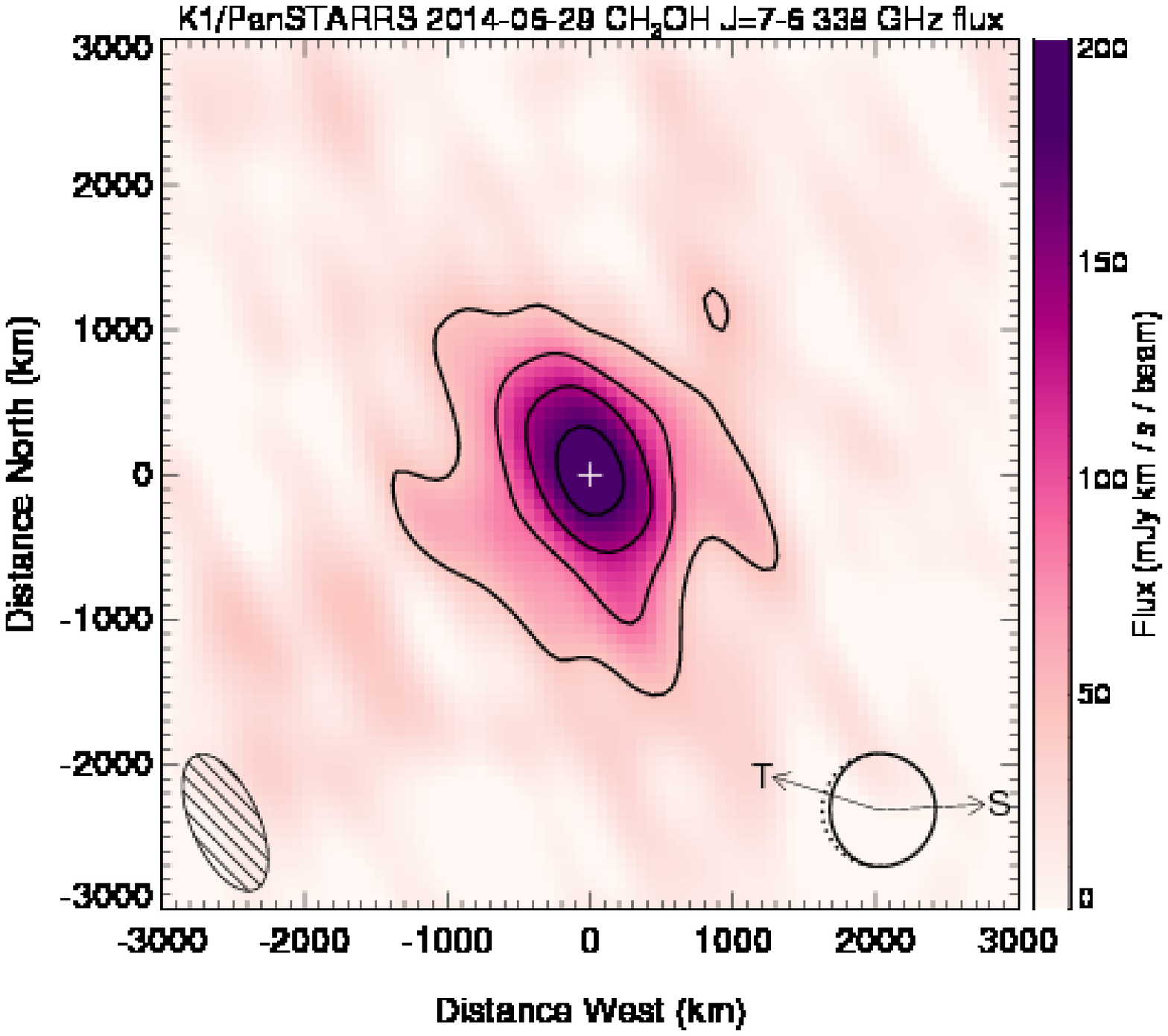}
\caption{ALMA flux maps of CH$_3$OH in comet C/2012 K1 (PanSTARRS) observed 2014-06-28 at 252~GHz (left) and 2014-06-29 at 338~GHz (right). White crosses indicate the emission peaks,  which are employed as the origin of the coordinate axes. Contours are plotted at $3\sigma$ intervals, where $\sigma$ is the RMS noise in each map. The PSF FWHM are shown lower left. Direction of Sun (S) and orbital trail (T) are indicated lower right along with the illumination phase.\label{fig:maps}}
\end{figure*}

\begin{figure*}
\centering
\includegraphics[width=0.45\textwidth]{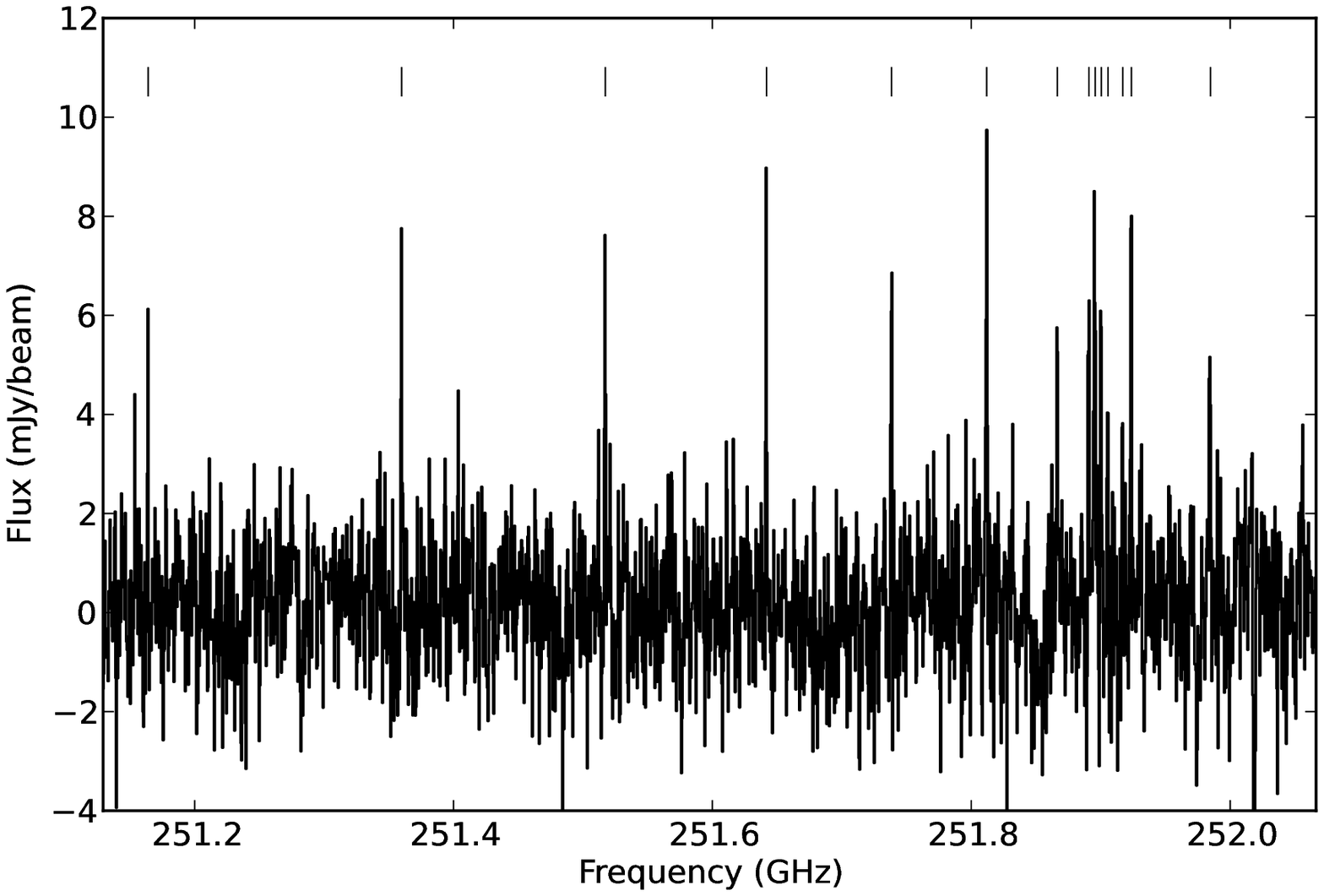}
\hspace{4mm}
\includegraphics[width=0.45\textwidth]{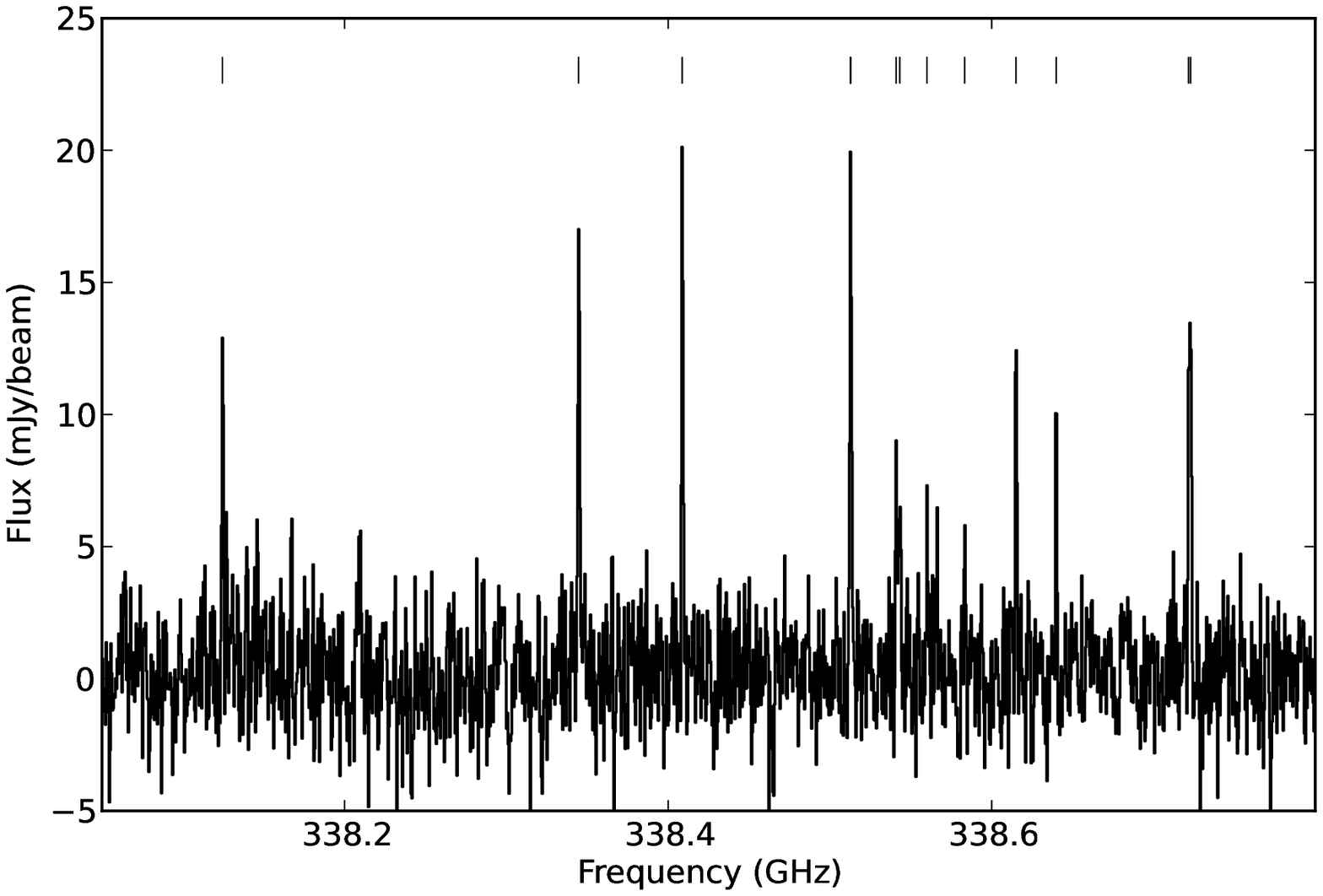}
\caption{CH$_3$OH spectra observed on 2014-06-28 of the $K=3-2$ band (left) and on 2014-06-29 of the $J=7-6$ band (right). Ticks indicate detected CH$_3$OH lines. These spectra were averaged over a $0.1''$-thick annulus centered on the nucleus, with an inner radius of $0.2''$\label{fig:spec}}
\end{figure*}

\begin{figure*}
\centering
\includegraphics[width=0.49\textwidth]{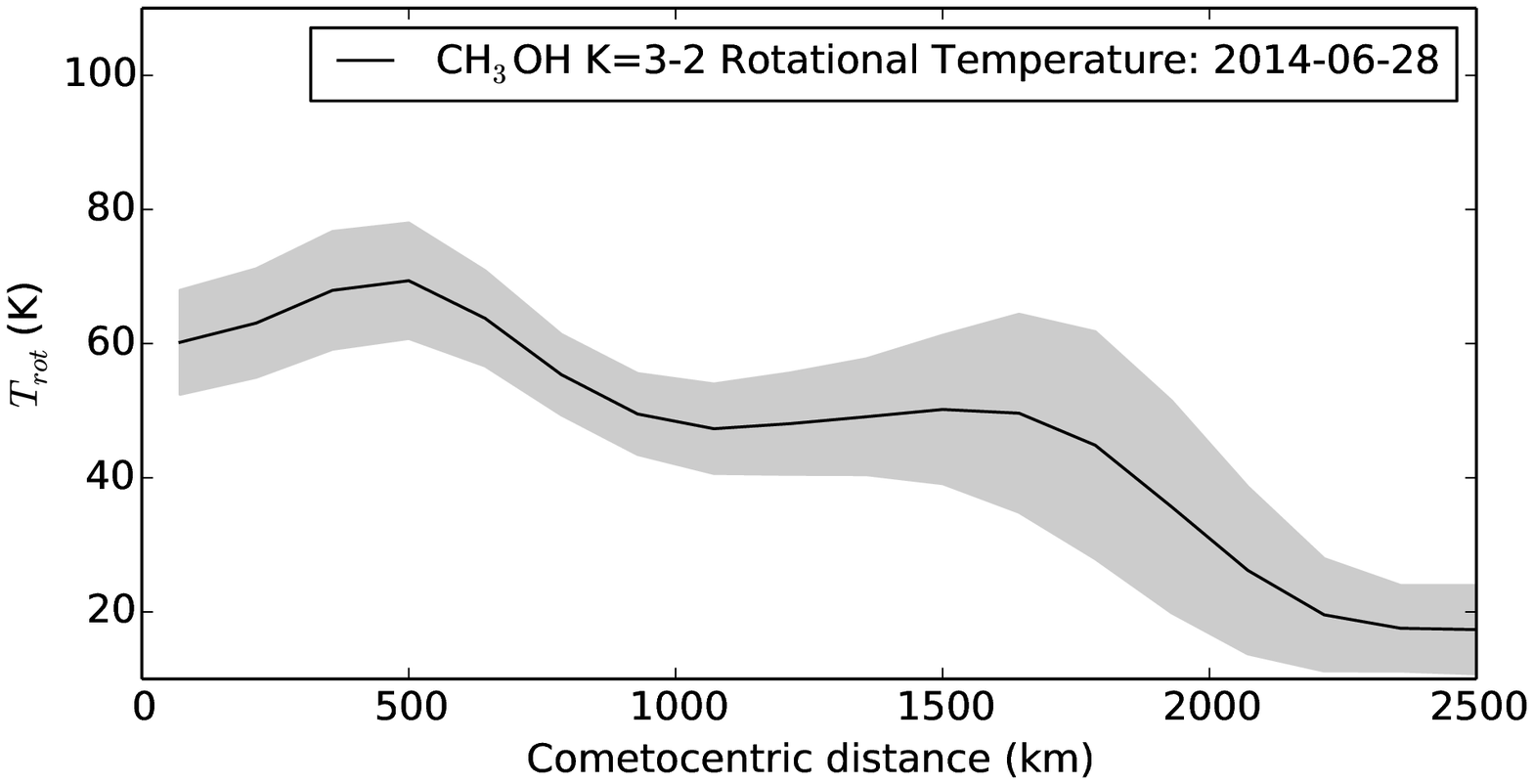}
\includegraphics[width=0.49\textwidth]{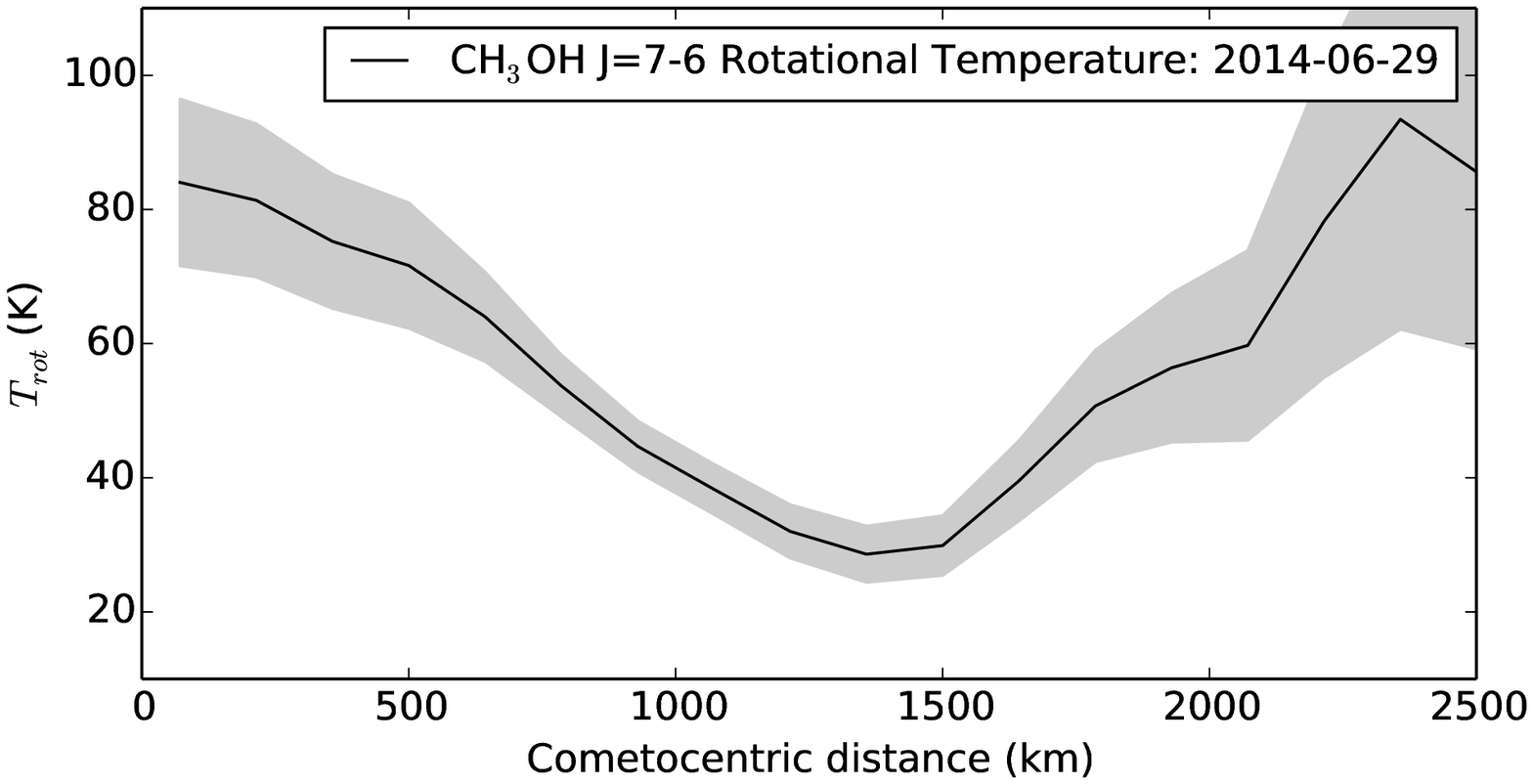}
\caption{Azimuthally-averaged temperature profiles for the $K=3-2$ (252~GHz) band (left) and the $J=7-6$ (338~GHz) band (right). Statistical $1\sigma$ error envelopes are shown. \label{fig:profiles}}
\end{figure*}

\section{Results}

Fig. \ref{fig:maps} shows the observed CH$_3$OH flux maps integrated over the strongest lines detected in each band. To first order (given the noise), these ALMA maps are consistent with azimuthal symmetry about the peak, and can be modeled under the assumption of an isotropic outflow from the nucleus \citep{cor14}. To obtain the highest spectroscopic signal-to-noise ratio as a function of radius, the spectral data cubes were azimuthally averaged about the emission peak, allowing for a significant improvement in sensitivity (particularly towards larger radii). The data were binned into a series of $0.1''$-thick annuli centered on the emission peak, and the average flux was taken in each annulus, resulting in azimuthally-averaged spectra ($\bar{S_{\nu}}$) as a function of sky-projected radius $\rho$ from the optocenter. The $\bar{S_{\nu}}$ for $\rho=0.25''$ ($\approx360$~km) are shown in Fig. \ref{fig:spec}.

From the observed emission line strengths, an excitation diagram technique was used to derive the CH$_3$OH rotational temperature ($T_{rot}$) as a function of $\rho$, following a similar method to that described by \citet{dis06}. Integrated line intensities (and their statistical errors) for each annulus were first converted to brightness temperature units (K\,km\,s$^{-1}$) using the Rayleigh-Jeans approximation. Then using the standard (optically thin) relationship between brightness temperature and column density \citep[\emph{e.g.}][]{cum86}, the column density ($N_i$) corresponding to each line flux ($S_idv$) was plotted as a function of upper-state energy ($E_i$), and the gradient ($m$) of this diagram was derived. Assuming LTE, the excitation temperature for which $m=0$ represents $T_{rot}$ (the mean rotational temperature along the line of sight), and the error on $T_{rot}$ is calculated from $\sigma_{m}dT_{rot}/dm$. This technique has the advantage over the conventional rotational diagram method that blended lines can be included through use of the weighted mean energy ($<{E_i}>$) of the blended transitions. Further details of this analysis will be provided in a future article (Cordiner et al., in prep.).

\section{Discussion and Conclusion}

Fig. \ref{fig:profiles} shows the azimuthally-averaged temperature profiles as a function of sky-projected distance from the emission peak, for both sets of CH$_3$OH observations. Both datasets show a general trend for decreasing rotational temperatures as a function of distance between $\rho=0$ and 1500~km, although the decrease was much slower on June 28th than June 29th. In fact, on June 28th (for the $K=3-2$ band), a possible increase in $T_{rot}$ occurred (from $60.2\pm7.8$~K to $69.4\pm8.6$~K) between $\rho=0$ and 500~km. This was followed by a relatively slow decrease with radius to $17.6\pm6.4$~K at $\rho=2500$~km. Conversely, on June 29th (for the $J=7-6$ band), there was a rapid decrease in $T_{rot}$ from $84.0\pm2.4$~K to $28.6\pm4.2$~K between $\rho=0$ and 1400~km, followed by a significant increase to $90\pm30$~K at $\rho\approx2400$~km. 

The maximum radial extent of the CH$_3$OH flux detected by ALMA was approx. $2500$~km. However, contemporaneous position-switched observations of the 338~GHz CH$_3$OH band were obtained using the APEX telescope on June 29th. These (single-pointing) data provide a measure of the mean rotational temperature within the $18''$ (26,000~km) APEX beam. A value of $T_{rot}=20\pm7$~K was derived, which shows that over the larger values of $\rho$ probed by APEX, the temperature was significantly less than in the inner few thousand km probed by ALMA. Such variations in temperature with distance from the nucleus have been observed previously by long-slit IR spectroscopy \citep[\emph{e.g.}][]{bon13}, and are predicted by theoretical studies due to adiabatic cooling of the outflowing gas \citep[see][and references therein]{rod05}. 

Interpretation of CH$_3$OH rotational temperatures is non-trivial due to the low density of the coma, as demonstrated by \citet{boc94}. Radiative cooling combined with decreased collision rates as a function of radius results in departure from local thermodynamic equilibrium (LTE), such that with increasing $\rho$, $T_{rot}$ falls increasingly below the coma kinetic temperature $T_{kin}$. Detailed excitation modelling (including radiative pumping, cooling and collisions between CH$_3$OH, H$_2$O and electrons), shows that $T_{rot}(J=7-6)$ falls more rapidly as a function of $\rho$ than $T_{rot}(K=3-2)$, due to the larger radiative relaxation rates of the $J=7$ levels. On both dates, however, non-LTE effects alone are insufficient to explain the observed temperature drops with increasing $\rho$, so these are attributed to adiabatic cooling of the coma. The increasing temperature with distance between 1500-2500~km on June 29th is difficult to explain, but a possible source of heating may be through the sublimation of icy grains in the coma, heated by Solar radiation \citep[see][]{fou12}.

As a caveat to our analysis, the ALMA observations are not sensitive to large-scale flux from the outer coma (on angular scales $\gtrsim5''$ or $\rho\gtrsim4000$~km), which could bias our derived $T_{rot}(\rho)$ curves. The magnitude of this effect will be quantified through detailed simulations in a future article, but the optimal method to deal with this problem will be through direct (line by line) modelling of the interferometric visibilities.

Our ALMA observations of CH$_3$OH emission at millimeter and submillimeter wavelengths have permitted the first `instantaneous', spatially-resolved 2-D measurements of rotational temperatures in the inner few thousand kilometers (sky-projected distance) of a cometary coma. Large variations in the CH$_3$OH rotational temperature in C/2012 K1 (Panstarrs) over distances $\sim1000$~km are likely due to variations in the coma kinetic temperature, the cause of which will be investigated in more detail in a future article. This study demonstrates that spatial temperature variations may need to be considered when deriving coma molecular abundances from spectral line data. Further high resolution observations and modelling are required in order to better understand the coma thermal physics and molecular excitation, and to assist in the determination of more accurate cometary compositions.

{\footnotesize This paper makes use of the following ALMA data:
ADS/JAO.ALMA\#2013.1.01061.S. ALMA is a partnership of ESO (representing
its member states), NSF (USA) and NINS (Japan), together with NRC
(Canada), NSC and ASIAA (Taiwan), and KASI (Republic of Korea), in
cooperation with the Republic of Chile. The Joint ALMA Observatory is
operated by ESO, AUI/NRAO and NAOJ.}

\end{document}